\documentclass[twocolumn,9pt]{article} % here we use the article class, rather than elsarticle

\usepackage[square,numbers,sort&compress,comma]{natbib}

\usepackage{amsmath}
\usepackage{amssymb}
\usepackage{caption}
\usepackage{graphicx}
\usepackage{latexsym}
\usepackage{times}
\usepackage[pagewise]{lineno}
\usepackage{hyperref}

\topmargin - 12pt % might need to be set to 0pt for some installations
\renewenvironment{abstract}%
              {% - begin definition
               \small% - select font
               {\bfseries \abstractname}% - select font
               \par% - end a paragraph (skip \parsep)
               \vspace{10pt}% - add vertical space
              }% - complete definition

\renewcommand\abstractname{Abstract}

\newcommand{\nomenclature}% - name of command
              [1]% - number of arguments
              {% - begin definition
               \bgroup% - begin a local group
               \flushleft% - turn on flushleft option
               \small\bf% - select font
               #1% - insert title text
               \par% - end a paragraph (skip \parsep)
               \egroup% - terminate local group
              }% - complete definition

\renewcommand{\section}% - name of command
              [1]% - number of arguments
              {% - begin definition
               \bgroup% - begin a local group
               \flushleft% - turn on flushleft option
               \small\bf% - select font
               \refstepcounter{section}% - increment counter
               \arabic{section}. #1% - insert title text
               \par% - end a paragraph (skip \parsep)
               \egroup% - terminate local group
              }% - complete definition

\renewcommand{\subsection}% - name of command
              [1]% - number of arguments
              {% - begin definition
               \bgroup% - begin a local group
               \flushleft% - turn on flushleft option
               \small\em% - select font
               \refstepcounter{subsection}% - increment counter
               \arabic{section}.% - insert title text
               \arabic{subsection}. #1% - insert title text
               \par% - end a paragraph (skip \parsep)
               \egroup% - terminate local group
              }% - complete definition

\renewcommand{\subsubsection}% - name of command
              [1]% - number of arguments
              {% - begin definition
               \bgroup% - begin a local group
               \flushleft% - turn on flushleft option
               \small\em% - select font
               \refstepcounter{subsubsection}% - increment counter
               \arabic{section}.% - insert title text
               \arabic{subsection}.% - insert title text
               \arabic{subsubsection}. #1% - insert title text
               \par% - end a paragraph (skip \parsep)
               \egroup% - terminate local group
              }% - complete definition

  \newcommand{\acknowledgement}% - name of command
              [1]% - number of arguments
              {% - begin definition
               \bgroup% - begin a local group
               \flushleft% - turn on flushleft option
               \small\bf% - select font
               #1% - insert title text
               \par% - end a paragraph (skip \parsep)
               \egroup% - terminate local group
              }% - complete definition

  \newcommand{\sectionbib}% - name of command
              [1]% - number of arguments
              {% - begin definition
               \bgroup% - begin a local group
               \flushleft% - turn on flushleft option
               \small\bf% - select font
               #1% - insert title text
               \par% - end a paragraph (skip \parsep)
               \egroup% - terminate local group
              }% - complete definition

\begin{document}

% -------------------------------------------------------------------- %
% -------------------------------------------------------------------- %
% -------------------------------------------------------------------- %

% -------------------------------------------------------------------- %

\small
\baselineskip 10pt

% -------------------------------------------------------------------- %

\thispagestyle{empty}
\twocolumn[\begin{@twocolumnfalse}

%% Title and Author Information for Editors and Evaluators 
\begin{center}
    \LARGE\bfseries Turbulent hydrogen premixed flames at high pressure and high temperature
\end{center}

\begin{center}
    \large Sofiane Al Kassar$^*$, Sara Cantagalli, William Lauder, Geveen Arumapperuma, Antonio Attili 
\end{center}

\begin{center}
    \footnotesize $^*$School of Engineering, Institute for Multiscale Thermofluids, The University of Edinburgh, Edinburgh, EH9 3FD, United Kingdom
\end{center}
% -------------------------------------------------------------------- %
%%% Note: The explanatory material in italic font on this page should be removed prior to manuscript submission.

{\bf 1) Novelty Statement}\\
This study presents the first systematic analysis of turbulent lean premixed hydrogen flames under simultaneous increases of pressure and temperature, to emulate isentropic compression in gas-turbine operations. Using Direct Numerical Simulations carefully scaled to preserve the same nominal turbulence levels, the configuration enables a direct comparison of turbulence–flame interactions across operating conditions. The results show that the coupled increase produces weaker overall changes than increasing pressure alone, but still leads to measurable modifications in flame structure and reactivity. These differences arise primarily from reduced turbulence dissipation within the flame, which sustains stronger turbulence and enhances thermodiffusive effects.
\vspace{10pt}

{\bf 2) Significance Statement}\\
The simultaneous increase of pressure and temperature, representative of gas-turbine operations, produces moderate overall changes but clear differences in flame structure and turbulence–flame coupling. This indicates that analyses and experiments performed at ambient conditions remain broadly relevant for extrapolation to gas-turbine operations. Reduced turbulence dissipation within the flame, due to smaller expansion and temperature ratios, sustains stronger turbulence and enhances thermodiffusive effects while maintaining the universal scaling of tangential strain with the Kolmogorov time at elevated conditions. The presented framework also enables the study of higher turbulence levels within the flame without additional computational cost.
\vspace{10pt}

{\bf 3) Author Contribution}\\
\begin{itemize}
  \item SAK: performed simulations, analysed data, wrote paper
  \item SC: performed simulations, analysed data
  \item WL: performed simulations, analysed data
  \item GA: code implementation
  \item AA: designed research, reviewed paper, supervision
\end{itemize}

\vspace{10pt}

{\bf 4) Preferred Presentation Format}\\
The authors prefer oral presentation for the following reasons:
\begin{itemize}
\item The results reveal new insights into turbulence–flame interactions and thermodiffusive effects, which are best communicated through an oral presentation with detailed images and videos, allowing the audience to be guided through the key physical mechanisms.
\item The presentation can focus directly on the main outcomes and insights, without requiring extensive background material, ensuring a clear and efficient discussion of the findings.
\item The findings are relevant to the hydrogen combustion and gas-turbine research communities, and are likely to stimulate engaging discussion with the audience.
\end{itemize}
\vspace{10pt}

{\bf 5) Colloquium/Colloquia Designation and Keywords}\\
Colloquium choices (order of preference): a) Turbulent flames; b) Flame dynamics and transport processes

   
\end{@twocolumnfalse}] 
\clearpage
% -------------------------------------------------------------------- %
% -------------------------------------------------------------------- %
% -------------------------------------------------------------------- %
\setcounter{page}{1}
% -------------------------------------------------------------------- %
\title{\LARGE \bf Turbulent hydrogen premixed flames at high pressure 
                  \\ and high temperature
                    }
                    
% \author{{\large Author 1 full name$^{a,*}$, Author 2 full name$^{a,b}$, Author 3 full name$^{b}$, $\ldots$}\\[10pt]
%         {\footnotesize \em $^a$Author affiliation 1}\\[-5pt]
%         {\footnotesize \em $^b$Author affiliation 2}\\[-5pt]
%         {\footnotesize \em Continue the list of affiliations as needed, with one per line}}

\author{
\parbox{\textwidth}{\centering
\large Sofiane Al Kassar$^{a*}$, Sara Cantagalli$^{b}$, William Lauder$^{a}$, 
        \\ Geveen Arumapperuma$^{a}$, Antonio Attili$^{a}$\\[10pt]
        {\footnotesize \em $^a$School of Engineering, Institute for Multiscale Thermofluids, The University of Edinburgh, Edinburgh, EH9 3FD, \\ United Kingdom}\\[-5pt]
        {\footnotesize \em $^b$Dipartimento di Ingegneria Civile e Industriale, Università di Pisa, 56122 Pisa, Italy}}}

\date{}  %%% Leave as is, do not add date;

% -------------------------------------------------------------------- %
% -------------------------------------------------------------------- %
% -------------------------------------------------------------------- %
\twocolumn[\begin{@twocolumnfalse}
\maketitle
\rule{\textwidth}{0.5pt}
\vspace{-5pt}

\begin{abstract} % 100 to 300 words.
%The influence of the combined increase of pressure and temperature, relevant to gas-turbine operations, on premixed lean hydrogen flames is investigated using Direct Numerical Simulations (DNS). The study focuses on the evolution of turbulence across the flame and its impact on the thermodiffusive effects. Three DNS are compared: one at ambient conditions ($1~\rm{atm}$ and  $298~\rm{K}$), one at intermediate ($5~\rm{atm}$ and  $472~\rm{K}$) and one at high pressure and temperature ($20~\rm{atm}$ and  $700~\rm{K}$). The three DNS are scaled to ensure the same jet Reynolds number $Re_{\rm jet}=11,200$ and the same nominal Karlovitz number in the unburnt side of the flame. Since pressure and temperature have opposing effects on the thermodiffusive instability, their combined influence results in minor differences in the laminar regime. However, despite negligible variations in one- and two-dimensional flames, the turbulent cases exhibit significant differences in flow-field structures and flame speed. The higher reactivity observed at elevated conditions is attributed to the lower expansion and temperature ratios across the flame, which reduce turbulence dissipation and therefore increase the impact of turbulence on the flame and strengthen thermodiffusive effects. Nevertheless
%
%Furthermore, it is shown that the universal scaling of tangential strain with Kolmogorov time observed in previous studies for homogeneous isotropic turbulence and a methane premixed flame also applies to hydrogen flames at different pressures and temperatures.
The combined influence of elevated pressure and temperature, representative of gas-turbine operating conditions, on lean premixed hydrogen flames is investigated using Direct Numerical Simulations (DNS) of a turbulent jet. Three cases are considered: 1 atm/298 K, 5 atm/472 K, and 20 atm/700 K, scaled to maintain the same jet Reynolds number and nominal Karlovitz number in the unburnt mixture, enabling a direct comparison of flame-turbulence interactions. Although the combined effects are moderate overall due to compensating influences, measurable differences arise in flame structure and turbulence-flame coupling. They are driven by reduced turbulence dissipation within the flame at high pressure and temperature, which enhances the interaction between turbulence and thermodiffusive effects. Finally, the tangential strain rate exhibits the same universal Kolmogorov scaling observed in homogeneous-isotropic turbulence and in methane flames, confirming its robustness for modelling turbulence.
\end{abstract}

\vspace{5pt}
\parbox{\textwidth}{\footnotesize {\em Keywords:} Hydrogen combustion; Direct numerical simulation; Thermodiffusive instability; Turbulent flames}
\rule{\textwidth}{0.5pt}
*Corresponding author.
\end{@twocolumnfalse}] 
\vspace{5pt}

% \linenumbers
\section{Introduction\label{sec:introduction}} \addvspace{8pt}

%Hydrogen has become a central focus of research and development in the transition toward low-emission energy systems, particularly for power generation applications. However, its specific properties compared to conventional carbon-based fuels pose significant challenges for practical implementation \citep{RASUL2022116326, en16031141, ma16206680}. 
%Hydrogen's high diffusivity and reactivity are at the origin of combustion instabilities observed in lean premixed flames. In particular, the thermodiffusive instability strongly affects hydrogen combustion dynamics, directly influencing the heat release rate and the flame speed \citep{berger_characteristic_2019}. 
%The influence of the thermodiffusive instability has been extensively studied in various laminar flame configurations, where it has been shown to significantly affect flame behaviour under conditions relevant to industrial applications \citep{berger_intrinsic_2022-1, berger_intrinsic_2022-2}. It was also demonstrated that thermodiffusive effects persisted in turbulent flames and could interact synergistically with turbulence, further enhancing heat release and flame consumption rates \citep{AHMED2021111586, berger_synergistic_2022, berger2024effects}. 

Hydrogen’s high diffusivity and reactivity make lean premixed flames particularly sensitive to thermodiffusive effects, which strongly affect the heat release rate and flame speed \citep{berger_characteristic_2019}.
The effects of the thermodiffusive instability have been extensively investigated in laminar flames, where they have been shown to significantly alter flame behaviour \citep{berger_intrinsic_2022-1, berger_intrinsic_2022-2}. These effects also persist in turbulent regimes, where they interact synergistically with turbulence, further enhancing heat release and flame consumption rates \citep{AHMED2021111586, berger_synergistic_2022, berger2024effects}.
Although several studies have investigated these interactions across a range of pressures and turbulence intensities, the combined influence of elevated pressure and temperature in turbulent flames, conditions representative of gas-turbine applications, has received less attention \citep{CHU20232129}. 
The present work addresses this gap by analysing three turbulent flames in which pressure and temperature are simultaneously increased, reproducing the isentropic compression characteristic of gas-turbine operation.
The three direct numerical simulations (DNS) are scaled to maintain the same jet Reynolds number and Karlovitz number on the unburnt side of the flame. Because the coupled increase in pressure and temperature also yields comparable levels of thermodiffusive effects in the laminar flames, this setup enables a direct comparison of flame-turbulence interactions and their influence on the thermodiffusive behaviour.

% preserve the ratio between the chemical and flow time scales, thereby maintaining similar turbulence intensities while allowing the assessment of potential modifications in flame-turbulence interactions arising from higher pressures and temperatures. The coupled increase in pressure and temperature also yields comparable levels of thermodiffusive effects across the cases, facilitating the analysis.

\begin{figure*}[h!]
    \centering
    \vspace{-0.4 in}
    \includegraphics[width=5.64 in]{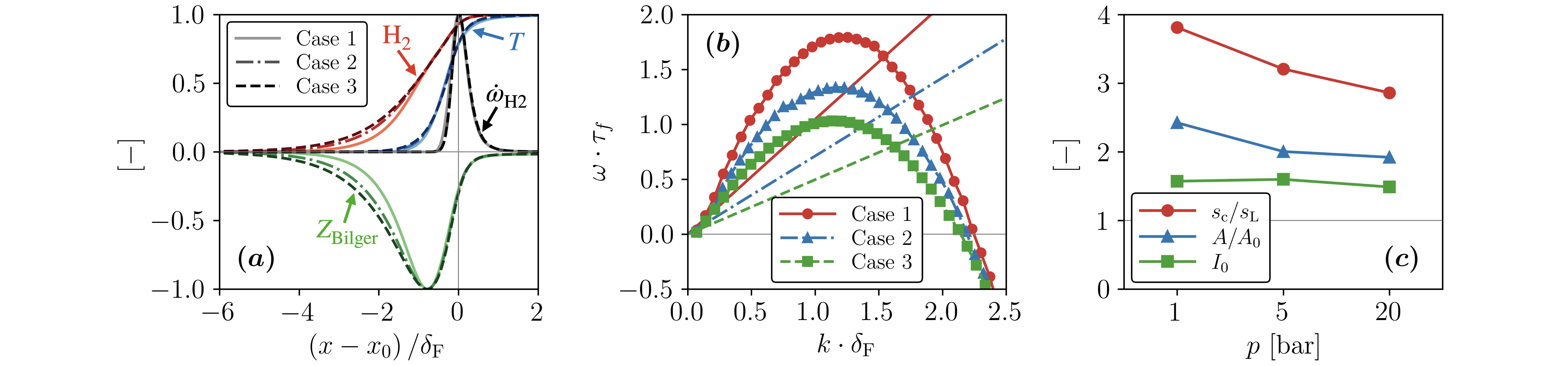}
    \includegraphics[width=5.64 in]{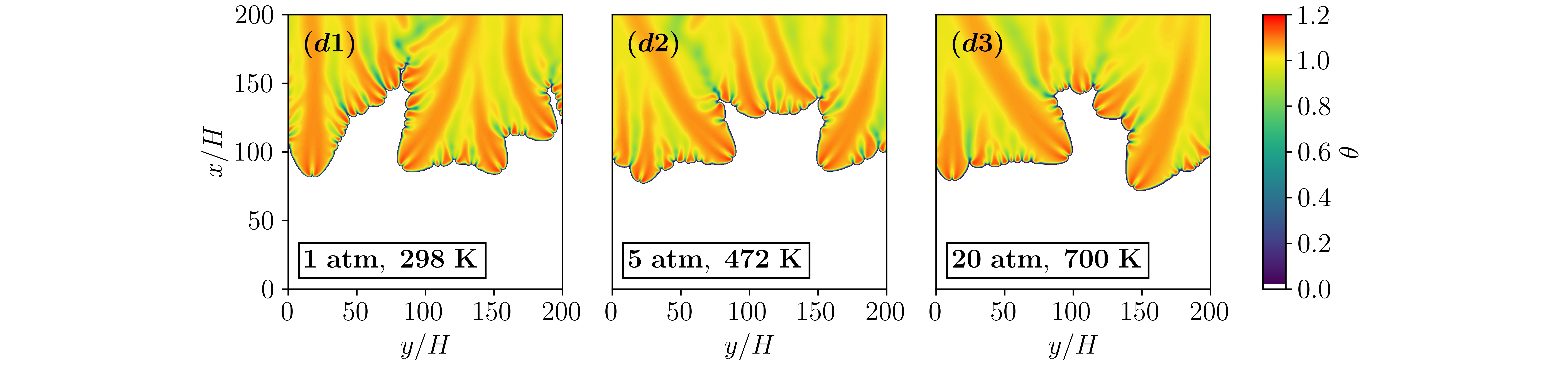}
    % \includegraphics[width=5 in]{figures/fields_2d_T.png}
    % \vspace{10 pt}
    \caption{\footnotesize Laminar references: (a) Normalised one-dimensional profiles (hydrogen mass fraction $\rm{H_2}$, temperature $T$, hydrogen reaction rate $\dot{\omega}_{\rm H_2}$ and Bilger mixture fraction $Z_{\rm Bilger}$), (b) dispersion relations, (c) non-linear regime values, (d.1-3) snapshots of progress variable for the different cases. The lines with no symbols in (b) correspond the Darrieus-Landau instability.}
    \label{fig:laminar_ref}
    \vspace{-3pt}
\end{figure*}

\section{Configuration, models, and methods} \addvspace{10pt}

%\subsection{Configuration and conditions for the 3D turbulent flames} \addvspace{10pt}

The configuration is a turbulent premixed hydrogen/air slot-jet flame operating at an equivalence ratio of $\Phi = 0.4$, similar to Ref.~\citep{berger_synergistic_2022, attili_turbulent_2021}. Three cases are considered, one corresponding to typical ambient conditions (Case 1: $p=1~\rm{atm}$, $T_u=298~\rm{K}$) and two under gas-turbine conditions at intermediate (Case 2: $p=5~\rm{atm}$ and $T_u=498~\rm{K}$) and high pressure and temperature (Case 3: $p=20~\rm{atm}$, $T_u=700~\rm{K}$).
The inlet bulk velocity $U_{\rm bulk}$ and the slot width $H$ are scaled with the
% the thermal flame thickness ($\delta_{\rm F}$) and 
kinematic viscosity $\nu$ to ensure the same jet Reynolds numbers $Re_{\rm jet} = UH/\nu=11,200$. Classical turbulence scalings for the Kolmogorov scale $\eta$ are employed to ensure that the nominal Karlovitz number $Ka$ remains the same in the unburnt mixture. 
The Kolmogorov scale is defined as $\eta=(\overline{\nu}^3/\tilde{\epsilon})^{1/4}$, where $\tilde{\epsilon}$ is the Favre-averaged energy dissipation and $\overline{\nu}=\overline{\mu/\rho}$ is the Reynolds-averaged viscosity, and $Ka=\delta_{\rm F}^2/\eta^2$, where $\delta_{\rm F}$ is the thermal flame thickness. The ranges of $\eta$, $Ka$, and Taylor microscale Reynolds number $Re_{\lambda}$ measured in the middle of the flame brush are summarised in Tab.~\ref{tab:sum_res}. The Taylor microscale Reynolds number is $Re_{\lambda}=2\tilde{k}\sqrt{5/(\overline{\nu}\tilde{\epsilon})}$ where $\tilde{k}$ is the Favre-averaged turbulent kinetic energy. The evolution of these quantities across the flame brush is analysed in Section~4.2.

\begin{table}[h] \footnotesize
    \caption{Conditions, one-dimensional characteristics, simulation parameters and turbulent characteristics of the three DNS cases.}
    \vspace{10pt}
    \centerline{\begin{tabular}{c|ccc}
    \hline 
    & Case 1&Case 2&Case 3\\
    \hline
    $p~{\rm [atm]}$&1 &5&20\\
    $T~{\rm [K]}$& 298 &472& 700\\
    $\phi~{\rm [-]}$& 0.4 &0.4& 0.4\\
    \hline 
    $s_{\rm L}~{\rm[m/s]}$& 0.186& 0.314&0.555\\
    $\delta_{\rm F}~{\rm [\mu m]}$& 707& 136&31.5\\
 $\sigma$& 4.44& 3.10&2.35\\
 $Le_{\rm eff}$& 0.34& 0.34&0.34\\
 $Ze$& 11.5& 12.0&11.4\\ 
    \hline
    $H~{\rm [mm]}$& 7.82 &1.55& 0.36\\
    $U_{\rm bulk}~{\rm [m/s]}$& 25.8 &57.0&120.3\\
    $U_{\rm coflow}~{\rm [m/s]}$& 5.17 &11.4& 24.1\\
    $Re_{\rm jet}$& 11200 &11200& 11200\\
    \hline
    $\Delta~{\rm [ \mu m]}$& 65 &13.0& 3.0\\
    % $L_x/H$& 12.5 &12.5 & 12.5\\
    % $L_y/H$& 15 &15 &15\\
    % $L_z/H$& 4.27 &4.27 &4.27\\
    % $N_x$& 1500 &1500 & 1500\\
    % $N_y$& 1800 &1800 &1800\\
    % $N_z$& 512 &512 &512\\
    % \hline
    $\delta_{\rm F}/\Delta$& 10.9 &10.5& 10.5\\
    $\eta_{\rm min}/\Delta$& 0.6&0.6& 0.6\\
    \hline
    $Re_{\lambda}$& 9-20& 16-26& 26-40\\
 $\eta/H\cdot1000$& 12-20& 9-15& 7-10\\
    $Ka$& 20-40&40-70& 70-100\\
    \end{tabular}}
    \label{tab:sum_res}
    \vspace{-3pt}
\end{table}

\begin{figure*}[h!]
    \centering
    \includegraphics[width=5.64 in]{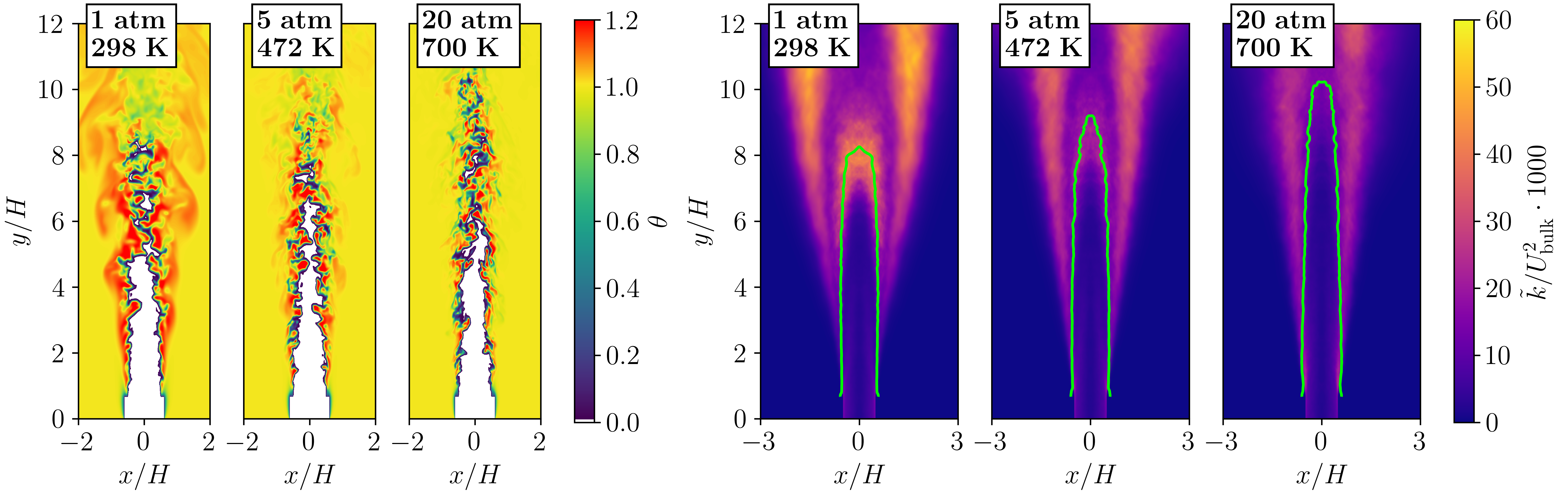}
    % \vspace{10 pt}
    \caption{\footnotesize Two-dimensional slices of the turbulent flames for the three cases: Progress variable based on temperature $\theta$ (left), turbulent kinetic energy $\tilde{k}$ normalised with the bulk velocity $U_{\rm bulk}$ (right). Isocontours of the progress variable based on hydrogen $C_{\rm H_2}=0.93$, corresponding to the peak reaction rate, are represented on top of $\tilde{k}$ to help locate the flame.}
    \label{fig:fields_3d}
\end{figure*}

%\subsection{Physical model and numerical methods} \addvspace{10pt}

The reactive, unsteady Navier-Stokes equations are solved under the low Mach number approximation. A finite-rate multistep chemistry model involving 9 species~\citep{burke_comprehensive_2012} is employed, with transport properties computed using a mixture-averaged model~\citep{attili_effects_2016}. The thermodiffusion (Soret) effect is included using the model proposed by Schlup and Blanquart~\cite{schlup_reduced_2018}. The governing equations are solved using a semi-implicit finite difference method~\citep{desjardins_high_2008}, extensively applied before in various configurations~\citep{attili_effects_2016, berger_characteristic_2019, attili_turbulent_2021, berger_synergistic_2022}. 
%Spatial derivatives are discretised using second-order finite differences for the momentum equation and scalar diffusive terms, while a third-order weighted essentially non-oscillatory (WENO) scheme~\citep{liu_weighted_1994} is employed for the convective terms in the scalar equations. 
%The domain is periodic in the spanwise direction ($z$), with open boundary conditions prescribed at the outlet in the streamwise direction ($y$), and slip conditions imposed at the lateral boundaries ($x$).
%The inlet velocity field is generated from an auxiliary simulation of a fully developed turbulent channel flow. 
A uniform mesh is used in all three directions, with a grid resolution $\Delta$ satisfying $\Delta/\eta\leq2$ and $\delta_{\rm F}/\Delta \approx 10$. 
The number of grid points is $[1500,~1800,~512]$ along $x$, $y$ and $z$ for a total of 1.4 billion points. The size of the domain is $[12H,~15H,~4.27H]$.
The chemical source term is treated using Strang's operator splitting \citep{strang_construction_1968}, with integration performed by the stiff ordinary differential equation (ODE) solver CVODE~\citep{hindmarsh_sundials_2005}.

\section{Laminar references and thermodiffusive behaviour in the laminar regime} \addvspace{10pt}

Laminar flames are simulated under the same conditions as the 3D cases in both 1D and 2D configurations to establish a reference behaviour. As shown in Fig.~\ref{fig:laminar_ref}, the three cases display very similar profiles in the laminar regime. The normalised profiles of species mass fractions, temperature and reaction rate almost overlap in the reference 1D flamelets, while the overall flame topology remains almost identical in the non-linear regime in the 2D laminar configuration. The differences observed in the dispersion relations can be attributed to the decrease in the expansion ratio $\sigma$ with increasing temperature and pressure, which weakens the Darrieus-Landau instability. However, when growth rates are measured relative to the Darrieus-Landau line, they exhibit a very similar trend, indicating that thermodiffusive effects remain essentially unchanged with increasing pressure and temperature. This observation is further supported by the decomposition of the flame speed $s_{\rm c}$, which indicates that $s_{\rm c}$ decreases with increasing pressure and temperature due to a lower flame surface area, whereas the stretched factor $I_0$ remains nearly constant. In summary, these results demonstrate that despite the variations in pressure and temperature, the three cases exhibit comparable thermodiffusive behaviour in the laminar regime, and that the observed differences are primarily driven by changes in $\sigma$.

\begin{figure}[h!]
    \centering
    \includegraphics[width=2.65 in]{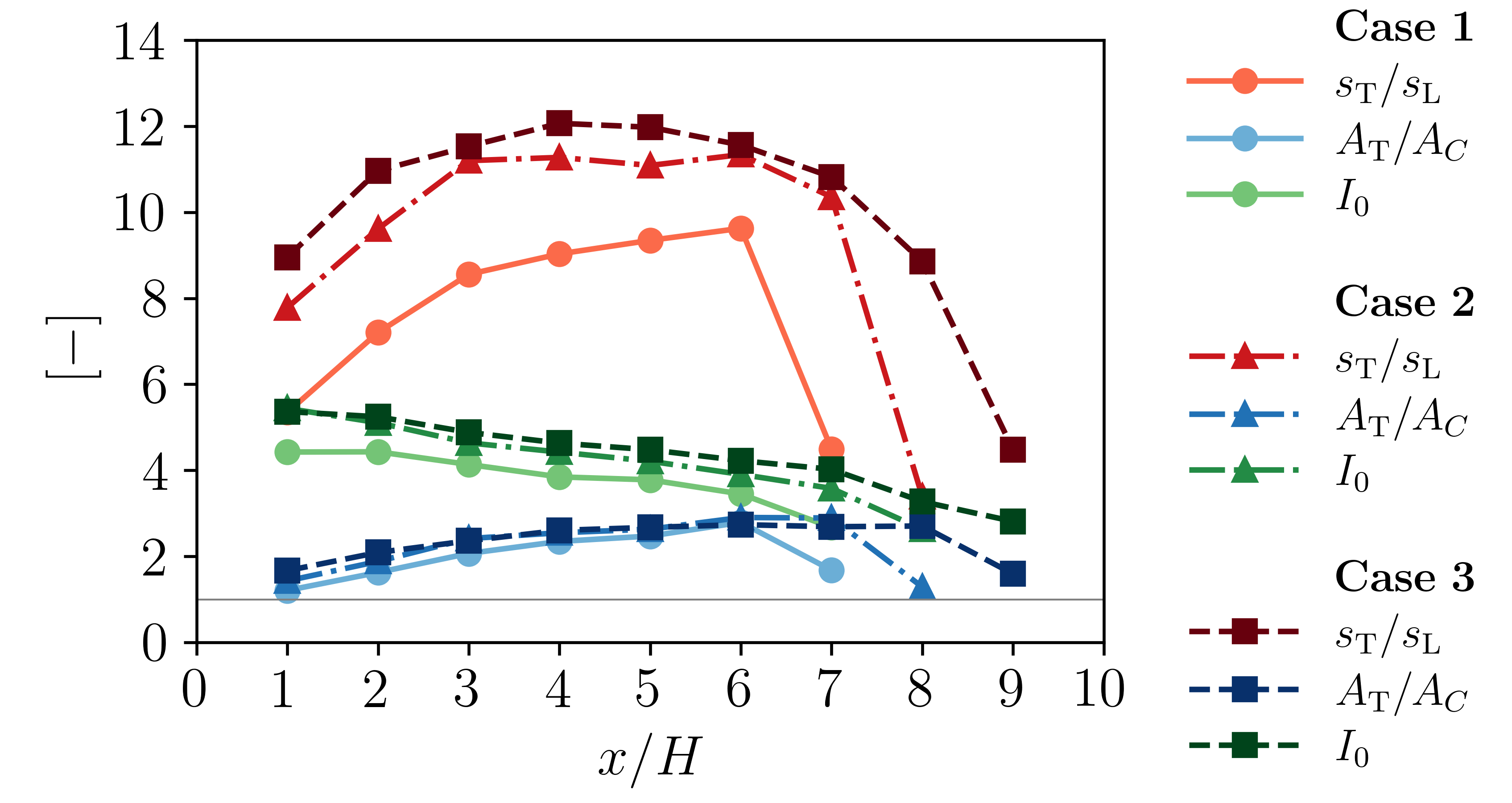}
    % \vspace{10 pt}
    \caption{\footnotesize Evolution of the turbulent flame speed $s_{\rm T}$, flame surface area $A_{\rm T}$ and stretched factor $I_0$ along the flame. $s_{\rm T}$ and $A_{\rm T}$ are respectively normalised with the laminar flame speed $s_{\rm L}$, and the reference flame surface $A_{C}$ obtained from an isocontour of mean progress variable.}
    \vspace{-3pt}
    \label{fig:st}
\end{figure}

\begin{figure*}[h!]
    \centering
,,    \includegraphics[width=5.64 in]{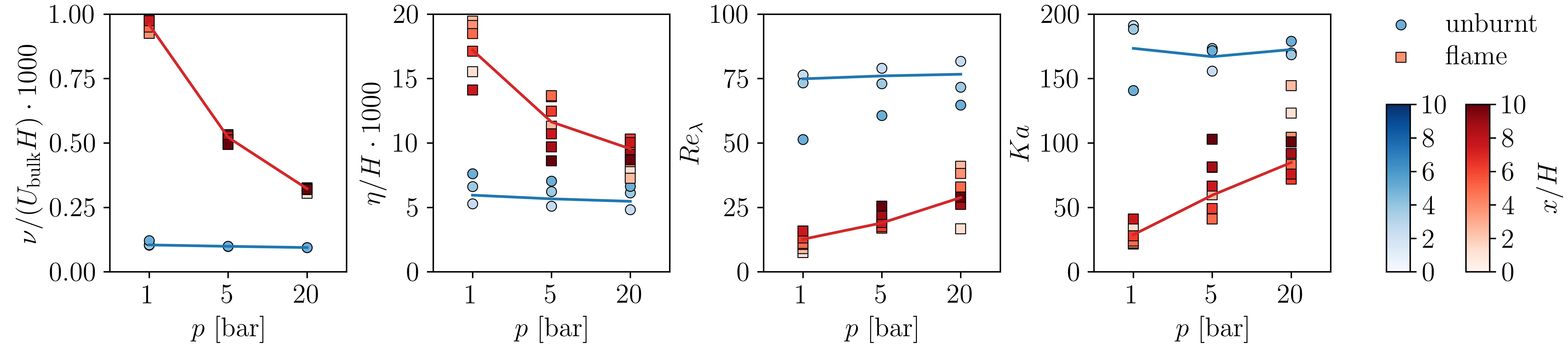}
    % \vspace{10 pt}
    \caption{\footnotesize Evolution of turbulent quantities with pressure and temperature at different streamwise locations in the unburnt jet and within the flame. From left to right: kinematic viscosity $\nu$, Kolmogorov scale $\eta$, Taylor microscale Reynolds number $Re_{\lambda}$ and Karlovitz number $Ka$. Blue circles and red squares represent values in the unburnt jet and within the flame, respectively. The colour shades become progressively darker downstream. Viscosity and Kolmogorov scale are normalised with the jet properties.}
    \label{fig:turb_prop}
\end{figure*}

\section{Results} \addvspace{8pt}
\subsection{Effect of high pressure and temperature on the turbulent flames} \addvspace{10pt}

%In contrast to the laminar cases, the differences between the three configurations become much more pronounced in the turbulent jets. As shown in Fig.~\ref{fig:fields_3d}, the flame length increases with pressure and temperature, accompanied by a clear change in the distribution of superadiatic temperatures. At ambient conditions, a thick layer of superadiabatic temperatures, also observed by \citet{berger_synergistic_2022}, forms along the outer part of the flame brush. At higher pressures and temperatures, however, the superadiabatic regions are confined to smaller, more localised structures. Significant differences are also observed in the turbulent kinetic energy ($\tilde{k}$). The bubble of turbulent kinetic production visible near the flame tip at ambient conditions, also reported by [have not found the ref in Lukas work and Day paper], disappears as pressure and temperature increase. This behaviour reflects a modification in the interactions between flame and turbulence at elevated conditions. This is also illustrated by the evolution of the flame speed shown in Fig.\ref{fig:st}. Unlike the 2D laminar cases, the flame speed increases with temperature and pressure. While the flame surface area is only weakly affected, the increase in $I_0$ demonstrates the enhancement of thermodiffusive effects at high pressure and temperature.

In the turbulent flames, the differences among the three cases remain moderate, as the simultaneous increase in pressure and temperature produces compensating effects. However, clear differences are still observed.
As shown in Fig.~\ref{fig:fields_3d}, increasing pressure and temperature markedly alters the instantaneous flow fields. At ambient conditions, a thick layer of superadiabatic temperatures, also reported by Berger et al.~\citep{berger_synergistic_2022}, develops along the outer edge of the flame brush. At higher pressures and temperatures, these regions become confined to smaller, more intermittent structures, indicating increased turbulent shredding of the high-temperature layer, with a stronger influence on the flame structure and surface. The turbulent kinetic energy field also exhibits significant differences when comparing the three cases. The region of high turbulent kinetic energy near the flame tip, clearly visible at ambient conditions and also reported in previous studies \citep{Chakraborty2011, berger_synergistic_2022}, disappears as pressure and temperature increase, indicating a substantial change in how turbulence is distributed across the flame.

The impact of the different turbulence-flame interactions is also reflected in the evolution of the turbulent flame speed, shown in Fig.~\ref{fig:st}. Unlike in the laminar cases, the turbulent flame speed increases with increasing pressure and temperature. While the flame surface area is only weakly affected, the higher values of $I_0$ confirm that thermodiffusive effects are amplified at elevated conditions due to the stronger coupling between turbulence and flame propagation.
Other quantities exhibit a similar trend, such as the local reaction rate and mixture fraction (not shown here for brevity), further confirming that thermodiffusive effects are increasingly enhanced by turbulence as pressure and temperature are raised. 

Overall, the differences between the cases remain moderate, indicating that analyses or experiments conducted at low pressure and temperature can still be extrapolated to gas-turbine conditions. Nonetheless, these variations are non-negligible and must be considered when developing or validating predictive models.

%[Need to work on this]
%Another important observation is the universality of the strain rate in these flames. The tangential strain ($K_{\rm S}$) is recognised to contribute significantly to surface generation in turbulent flames [Lucca, Berger] and the enhancement of thermodiffusive effects and reaction rate in turbulent hydrogen flames [Berger]. $K_{\rm S}$ is also used to improve tabulated models for turbulent hydrogen flames [Bottler, Porcarelli].  As shown in Fig.\ref{fig:st}, the strain rate normalised by the Kolmogorov time ($\tau_{\eta} = \overline{\nu}/\tilde{\epsilon}$) remains nearly constant along the flame, independently of pressure and temperature, demonstrating the same universal scaling observed for methane at different flames $Re$ [Lucca] and hydrogen flames at ambient conditions for different $Ka$. This confirms the robustness of the scaling law and its relevance for modelling.

%\textbf{NEED TO SAY SOMEWHERE THAT THE EFFECTS ARE IN GENERAL SMALL}

\subsection{Flame-turbulence interaction} \label{sec:flame_turb} \addvspace{10pt}

Fig.~\ref{fig:turb_prop} shows the evolution of the turbulent properties from the unburnt side to the middle of the flame brush. The normalised $\eta$, $Re_{\lambda}$ and $Ka$ remain the same in the unburnt mixture, confirming that the scaling employed to design the three cases is correct. As expected, $\eta$ increases across the flame while $Re_{\lambda}$ decreases, indicating that the flame is dissipating turbulence. The rate of this dissipation, however, depends strongly on the operating conditions. 
As pressure and temperature increase, $\eta$ decreases within the flame, while $Re_{\lambda}$ increases, indicating that smaller turbulent structures persist and that turbulence is less attenuated. This behaviour is due to the lower expansion ratios and temperature increase at elevated conditions, which weaken dilatation effects and increase in kinematic viscosity, responsible for turbulence dissipation. As a result, the Karlovitz number in the flame increases with pressure and temperature, even though it remains constant in the central jet. These higher $Ka$ explain the higher values of reaction rates and $I_0$ observed at elevated conditions, in line with the findings reported by Berger et al.~\citep{berger2024effects}. Despite the thermodiffusive effects being similar in the laminar configurations, they are clearly enhanced by turbulence at higher pressure and temperature. 

In summary, higher pressure and temperature reduce the flame's ability to dissipate turbulence, strengthening turbulence within the flame and reinforcing its impact on the flame. Interestingly, it also means that increasing pressure and temperature allows stronger flame-turbulence interaction while maintaining the same nominal level of turbulence in the unburnt mixture. This makes high-pressure, high-temperature cases particularly suitable for DNS studies, as they effectively achieve higher turbulence levels within the flame without increasing computational cost, while also maintaining similar thermodiffusive behaviour in 2D.

\subsection{Universality of tangential strain} \addvspace{10pt}

\begin{figure}[h!]
    \centering
    \includegraphics[width=2.65 in]{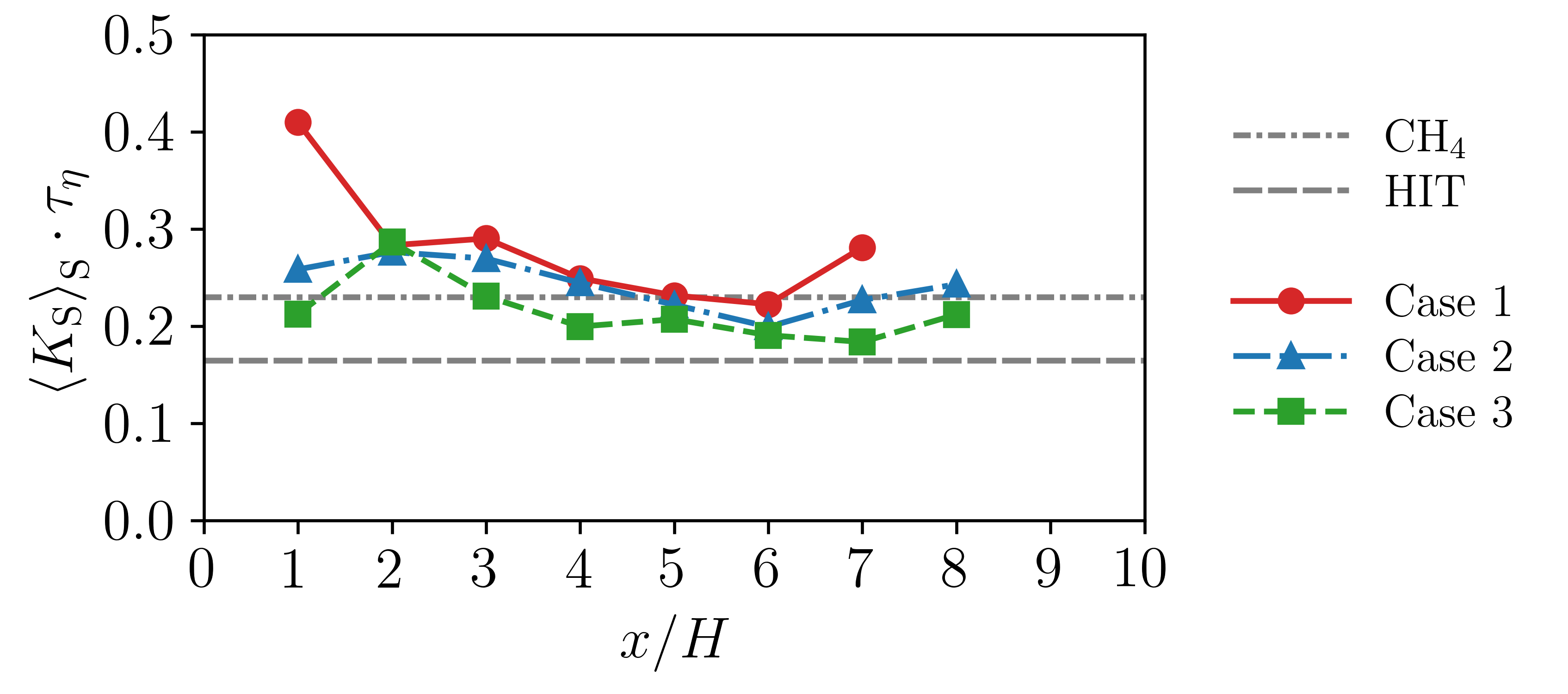}
    % \vspace{10 pt}
    \caption{\footnotesize 
    Evolution of tangential strain rate $\left\langle K_{\rm S}\right\rangle _{\rm S}$ inside the flame at different streamwise locations. $\left\langle K_{\rm S}\right\rangle _{\rm S}$ is the surface-weighted average of tangential strain conditioned on $C_{\rm H_2}=0.8$ and is normalised with the Kolmogorov time $\tau_{\eta}$. Universal strains obtained for homogeneous isotropic turbulence ($\rm{HIT}$) ~\citep{Girimaji1990427} and methane (${\rm {CH_4}}$)~\citep{LUCA20192451} are also represented.}
    \label{fig:ks}
    \vspace{-3pt}
\end{figure}

In turbulent flames, the relationship between turbulence and the enhancement of thermodiffusive effects is often associated with strain and the Karlovitz number \cite{berger2024effects}. Strain is also a key quantity in combustion modelling, where it is employed for reaction-rate parametrisation in tabulated and reduced-order models \cite{BOTTLER20241397, porcarelli2025assessment}.
Of particular interest is the component of the strain tangential to scalar surfaces, such as the flame front, which exhibits remarkable universality when normalised with Kolmogorov variables. Across a wide range of conditions, the normalised tangential strain attains a nearly constant value very close to 0.2, being about 0.165 in homogeneous isotropic turbulence \cite{Girimaji1990427} and approximately 0.23 in premixed methane and hydrogen flames across different Reynolds and Karlovitz numbers \cite{LUCA20192451, berger2024effects}. It is therefore of interest to assess whether this universal scaling also applies to hydrogen flames under elevated pressure and temperature. As shown in Fig.~\ref{fig:ks}, the strain rate normalised by the Kolmogorov time scale, $\tau_{\eta} = \overline{\nu}/\tilde{\epsilon}$, remains very close to the value 0.23 along the flame, independently of pressure and temperature, demonstrating the same universal scaling observed for homogeneous isotropic turbulence and for methane and hydrogen flames at ambient conditions. This confirms that the universal behaviour of the tangential strain is preserved under gas-turbine-relevant conditions, reinforcing its robustness and importance for turbulence-flame interaction modelling.

\section{Conclusion} \addvspace{10pt}

%The effects of the simultaneous increase in pressure and temperature, representative of gas-turbine conditions, on turbulent hydrogen premixed flames were studied through a series of DNS scaled to maintain the same nominal $Re$ and $Ka$ in the unburnt gases. A preliminary analysis in the laminar regime confirmed that the three configurations exhibited similar thermodiffusive behaviour. In contrast,  the turbulent flames exhibited clear differences: flame length, flame speed and $I_0$ increased with higher pressure and temperature, while superadiabatic regions became more localised. These observations are attributed to the reduced turbulence dissipation at elevated conditions, caused by the lower expansion and temperature ratios, which reinforced flame-turbulence interactions and enhanced thermodiffusive effects. 
%These simulations also confirmed that the universal scaling of the tangential strain with the Kolmogorov time, previously observed in homogeneous isotropic turbulence and in various methane flame configurations, also applies to hydrogen flames. Finally, it is worth noting that even under identical inlet turbulence conditions, elevated pressure and temperature can lead to higher turbulence intensity within the flame, providing a valuable opportunity to study stronger flame–turbulence interactions in similar configurations at no additional cost.

The simultaneous increase of pressure and temperature, representative of gas-turbine conditions, was investigated using DNS of turbulent lean premixed hydrogen flames. The three cases were scaled to maintain the same Reynolds and Karlovitz numbers in the unburnt mixture, allowing a direct comparison of turbulence–flame interactions across conditions. The results show that the combined effects of pressure and temperature remain moderate overall, as their opposing influences partly compensate. Nevertheless, measurable differences arise due to reduced turbulence dissipation within the flame at high pressure and temperature, caused by smaller expansion and temperature ratios. This sustains stronger turbulence inside the flame, enhancing thermodiffusive effects and reactivity, and also meaning that higher levels of turbulence can be achieved in the flame with no additional computational costs. The tangential strain rate retains its universal Kolmogorov scaling, confirming its robustness for combustion modelling.

\acknowledgement{Declaration of competing interest} \addvspace{10pt}
The authors declare that they have no known competing financial interests or personal relationships that could have appeared to influence the work reported in this paper.

%\acknowledgement{Acknowledgments} \addvspace{10pt}
%Archer2 blabla

% -------------------------------------------------------------------- %
% -------------------------------------------------------------------- %
% -------------------------------------------------------------------- %
\footnotesize
\baselineskip 9pt

% -------------------------------------------------------------------- %
% -------------------------------------------------------------------- %
% -------------------------------------------------------------------- %
\clearpage
\thispagestyle{empty}
\bibliographystyle{pci}
\bibliography{PCI_LaTeX}

% -------------------------------------------------------------------- %
% -------------------------------------------------------------------- %
% -------------------------------------------------------------------- %

\newpage

\small
\baselineskip 10pt

% -------------------------------------------------------------------- %
% -------------------------------------------------------------------- %
% -------------------------------------------------------------------- %

\end{document}